\documentclass{revtex4}
\usepackage{amssymb}
\usepackage{amsmath}

\input{tcilatex}

\begin{document}

\title{Dilatonic Entropic Force }
\author{I.Sakalli$^{\ast }$}
\affiliation{Department of Physics, Eastern Mediterranean University,}
\affiliation{G. Magusa, north Cyprus, Mersin-10, Turkey}
\affiliation{$^{\ast }$izzet.sakalli@emu.edu.tr}

\begin{abstract}
We show in detail that the entropic force of the static spherically
symmetric spacetimes with unusual asymptotics can be calculated through the
Verlinde's arguments. We introduce three different holographic screen
candidates, which are first employed thoroughly by Myung and Kim [Phys. Rev.
D 81, 105012 (2010)] for Schwarzschild black hole solutions, in order to
identify the entropic force arising between a charged dilaton black hole and
a test particle. The significance of the dilaton parameter on the entropic
force is highlighted, and shown graphically.
\end{abstract}

\maketitle

\section{Introduction}

Since Einstein's epochal studies \cite{Einstein} on general relativity (GR),
it has been argued that gravity is different at a deep level than other
fundamental forces. According to the GR, one should think of gravity as a
deformation of spacetime due to the presence of energy or stress. Although,
in the appropriate limit, GR reduces to Newtonian gravity, till the present
day there is no confirmed evidence for the gravitational fields, and their
source -- graviton. Because of this, whenever the agenda is about gravity,
new theories continue to attract much attention in physics. Recently,
Verlinde \cite{Verlinde} has invented a conceptual theory that gravity is
emergent, rather than the fundamental force. Essentially, Verlinde showed
that one can start from very general considerations, and gravity emerges as
an entropic force. In addition to this, Verlinde's arguments reveal a fact
that the key to understanding gravity is information (or entropy).

It is obvious that Verlinde was highly inspired by early works on
information storage in black holes by Bekenstein \cite{Bekenstein}, Hawking 
\cite{Hawking} and 't Hooft \cite{Hooft}. In the framework of quantum
gravity, such an information storage is explained best by the holographic
principle \cite{Hooft,Holography}, which was originally proposed by 't Hooft 
\cite{Hooft}. In fact, the holographic principle states that the entropy of
a black hole (BH) is proportional to the surface area of its event horizon
and not its volume; that volume itself is illusory and the universe is
really a hologram which is isomorphic to the information inscribed on the
surface of its boundary. Namely, the amount of information needed to
describe a BH can be entirely coded on its event horizon. Verlinde,
considering both holographic principle and the second law of thermodynamics 
\cite{Bekenstein}, showed that an entropic force is an apparent force that
results from changes in the information associated with the positions of
particles such that it drags particles towards and across the event horizon.
Videlicet, entropic forces are at work to pull particles across the
holographic screen. Macroscopically, this effect is nothing but the gravity.
Meanwhile, it should be noted that, prior to the Verlinde's work,
Padmanabhan \cite{Padmanabhan} also obtained the Newton's law of gravity by
using the equipartition law of energy; $E=2TS$. However, in addition to the
Newton's second law and Newtonian gravity, the astounding contribution
coming from Verlinde is the derivation of the Einstein's equations. Soon
after Verlinde's new proposal about the gravity, many relevant works
appeared. Some related (and very recently published) works can be seen in 
\cite{EFstudies}, and references therein.

In this paper, we think that it is interesting to consider the derivation of
Verlinde's entropic force in a spacetime with unusual asymptotics. To this
end, we would like to contribute to the literature, and try to learn more
about the concept of entropic force. Because, we believe that only when more
and more results on entropic force are available, Verlinde's arguments will
be more reliable. The paradigm of spacetimes with unusual asymptotics is
best described by the metric of charged dilaton black holes (CDBHs) \cite%
{CDBH}. In this paper we shall be concerned with CDBHs, which are stationary
and static. CDBHs have both asymptotically flat (AF) and non-asymptotically
flat (NAF) features depending on their dilaton parameter -- "$a$". The
beauty of CDBHs is that they have Schwarzschild limit when the dilaton field
vanishes. Here, we introduce three different candidates in order to define
the holographic screen (HS), locating at equipotential surfaces, of a CDBH:
accelerating surface (AS), static HS and stretched horizon (SH), which might
also be described as a Rindler spacetime. After giving the brief summary of
formulation of the entropic force, we calculate the each temperature of
these holographic screens, and plug the temperatures separately into the
derived entropic force formula. In doing this, we also verify that the
obtained entropic forces reduce to the results found in a very recent study 
\cite{Myung}, by Myung and Kim, in which the Schwarzschild spacetime is
considered, and thus to the Newton's force law. Of course, the ultimate aim
here is to highlight the effect of the dilaton parameter on the entropic
force by making use of numerical computation and plotting.

The outline of the present paper is as follows. A brief summary of the idea
underlying Verlinde's entropic force is given in the next section. Sec. III
and Sec. IV are devoted to the derivation of the entropic force by using AS
and static HS methods, respectively. Sec. V, we calculate the entropic force
on the SH. The paper ends with a conclusion, which appears in Sec. VI.

Throughout the paper, the natural units of $c=G=k_{B}=\hbar =1$ are used.

\section{A Brief Summary of Entropic Force}

Here we briefly mention some key points of the entropic force by following 
\cite{Verlinde}. As a demonstration to the usage of Verlinde's entropic
force, we retrieve the Newton's law of gravitation by using the holographic
screen in the non-relativistic case.

Verlinde postulated that the change of entropy on a holographic screen in
which a test particle of mass $m$ is located very close to it is linear with
distance $\Delta x$, namely

\begin{equation}
\Delta S=2\pi m\Delta x,
\end{equation}

According to the first law of thermodynamics, the entropic force exerting on
the test particle due to the change in entropy is

\begin{equation}
F\Delta x=T\Delta S,
\end{equation}

where $T$ is the temperature. After manipulating Eqs. (1) and (2), it is
easy to see that

\begin{equation}
F=2\pi mT,
\end{equation}

From here on, above identity will be our master equation to obtain the
entropic force. Simply put, once the temperature on the holographic screen
is defined, we just insert it into Eq. (3) and get the entropic force.

Remarkably, let us recall the equipartition rule of energy:

\begin{equation}
E=\frac{1}{2}\tilde{N}T,
\end{equation}

where $\tilde{N}$ is the partition number. As proposed by Verlinde, if we
combine the holographic principle and the equipartition rule of energy, $%
\tilde{N}$ is interpreted as the number of bits on the holographic screen.
In fact, this number is proportional to the area of the holographic screen as

\begin{equation}
\tilde{N}=cA,
\end{equation}

where $c$ is an constant, which is equal to one for AF and spherically
symmetric spacetimes. As it can be seen in the latter section, $c$ can take
values different than one depending on the non-asymptotic flatness of the
spacetime.

In order to reach the Newtonian force law, it is plausible to assume the
holographic screen as a spherical screen with radius $R$ which belongs to an
AF spacetime, whose its total mass $M$ is located at its centre. Thus, the
equipartition of energy rule takes the following form

\begin{eqnarray}
E &=&M=\frac{1}{2}AT,  \notag \\
&=&2ST,
\end{eqnarray}

where $S=\frac{A}{4}$ is the entropy of the holographic screen with the area 
$A=4\pi R^{2}.$ It is needless to say that $E$ is identified with the mass $%
M $ inside the screen. By using Eq. (6), one can easily read the temperature
on the holographic screen as

\begin{equation}
T=\frac{M}{2\pi R^{2}},
\end{equation}

As we emphasized before, once the temperature of the holographic screen is
determined, the entropic force is easily read from Eq. (3). Thus, after
plugging Eq. (7) into Eq. (3), we obtain the entropic force as

\begin{equation}
F_{N}=\frac{Mm}{R^{2}}.
\end{equation}

which is the well-known Newton's law of gravitation.

\section{Features of CDBH}

In the low-energy limit of string field theory, the four-dimensional action
(in Einstein frame) describing the dilaton field $\phi $ with its parameter $%
a$ (without loss of generality, we consider $a>0$)\ coupled to a $U(1)$
gauge field \cite{GHS} is

\begin{equation}
S=\int d^{4}x\sqrt{-g}(\Re -2(\nabla \phi )^{2}-e^{-2a\phi }F^{2})
\end{equation}

where $\Re $ is the curvature scalar and $F^{2}=F_{\mu \upsilon }F^{\mu
\upsilon }$ in which $F_{\mu \upsilon }$ is the Maxwell field associated
with a $U(1)$ subgroup of $E_{8}\times E_{8}$ or Spin(32)/$Z_{2}$. In the
presence of a electric charge with the non-constant the dilaton field, the
static spherically symmetric solutions are designated with CDBHs \cite{CDBH}
whose metric is given by

\begin{equation}
ds^{2}=-fdt^{2}+\frac{dr^{2}}{f}+R^{2}(d\theta ^{2}+\sin ^{2}\theta d\varphi
^{2}),
\end{equation}

where the metric functions $f$ and $R$ are

\begin{equation}
f=\frac{1}{\gamma ^{2}}r^{\frac{2}{1+a^{2}}}(1-\frac{r_{+}}{r}),
\end{equation}

and

\begin{equation}
R=\gamma r^{N},
\end{equation}

Here $r_{+}$ denotes the event horizon of the CDBHs, $\gamma $\ is an
arbitrary real constant, and another real constant $N$ is governed by

\begin{equation}
N=\frac{a^{2}}{1+a^{2}},
\end{equation}

The dilaton is found in the following form

\begin{equation}
\phi =\phi _{0}+\phi _{1}\ln r,
\end{equation}

where

\begin{equation}
\phi _{0}=-\frac{1}{2a}\ln \left[ \frac{Q^{2}\left( 1+a^{2}\right) }{\gamma
^{2}}\right] \text{ and }\phi _{1}=\frac{N}{a},
\end{equation}

where $Q$ refers to the electric charge. In this case, the solution for the
electromagnetic field is found as

\begin{equation}
F_{tr}=\frac{Qe^{2a\phi }}{R^{2}},
\end{equation}

CDBHs are not vacuum solutions since the action (9) contains a static
dilaton fluid whose energy-momentum tensor never vanishes. Essentially,
taking account of such a particular fluid model makes the CDBHs so special
that they are neither AF nor NAF.

By following the quasilocal mass definition for the NAF BHs, which was
defined by Brown and York \cite{BrownYork}, one can see that the horizon $%
r_{+}$ is related to the mass $M$ through

\begin{equation}
r_{+}=\frac{2M}{N}.
\end{equation}

For $r_{+}>0$, the horizon at $r=r_{+}$ hides the singularity located at $%
r=0 $. In the extreme case $r_{+}=0,$\ metric (10) still exhibits the
features of the BHs. Because the singularity at $r=0$ is null and marginally
trapped such that it prevents the outgoing signals to reach the external
observers. Besides all these, a CDBH has no inner horizon and no extremal
limit on the charge $Q$. This is one of the characteristics of the CDBHs.
Metric functions (11) and (12) correspond to the linear dilaton BHs \cite%
{LDBH} in the case of $a=1$. On the other hand, while $a\rightarrow \infty $
with $\gamma =1,$ metric (10) reduces to the Schwarzschild BH. Finally, it
is worth to note that the CDBHs have no zero charge limit.

\section{Accelerating Surface of CDBH and the Entropic Force}

The aim of this section is to define the entropic force on the AS of the
CDBH. The definition of the AS or so-called bulk is first introduced by M%
\"{a}kel\"{a} \cite{Makela}, and it has been recently used in \cite{Myung}
in order to derive the entropic force appeared in the Schwarzschild BH.
Here, we also consider the AS as if a HS, and we shall derive the entropic
force for the CDBHs.

If one defines a future pointing unit vector $u^{\alpha }$, which is the
congruence for the timelike world lines of the points on a spacelike
hypersurface $S^{2}$, it should satisfy the following orthogonality condition

\begin{equation}
u^{\alpha }n_{\alpha }=0,
\end{equation}

where $n_{\alpha }=[0,\frac{1}{\sqrt{f}},0,0]$\ is a normal vector on $S^{2}$%
. Generic choice of the future pointing unit vector is $u^{\alpha }=[\frac{1%
}{\sqrt{f}},0,0,0]$. Essentially, the reason of using the future pointing
unit vector $u^{\alpha }$ is to find the change of the heat, which is
related with the acceleration. Thus, it is plausible to introduce the local
acceleration in terms of $u^{\alpha }$\ as follows:

\begin{equation}
\tilde{a}^{\mu }=u^{\alpha }u_{;\alpha }^{\mu },
\end{equation}

where ";" denotes the covariant derivative. This implies that the only
nonzero component of the proper acceleration vector is

\begin{equation*}
\tilde{a}^{r}=\frac{f^{\prime }}{2}=\frac{1}{\gamma ^{2}(1+a^{2})}r^{\frac{%
1-a^{2}}{1+a^{2}}}\left[ 1+\frac{r_{+}(a^{2}-1)}{2r}\right] ,
\end{equation*}

in which a prime symbol denotes derivative with respect to $r$. The proper
acceleration is defined by

\begin{equation}
\tilde{a}=\tilde{a}^{\mu }n_{\mu },
\end{equation}

which yields

\begin{equation}
\tilde{a}=\frac{r^{-N}}{\gamma (1+a^{2})}\frac{\left[ 1+\frac{r_{+}(a^{2}-1)%
}{2r}\right] }{\sqrt{1-\frac{r_{+}}{r}}},
\end{equation}

By the virtue of the above result, one can read the Unruh temperature \cite%
{Unruh} as

\begin{equation}
T_{U}=\frac{\tilde{a}}{2\pi }=\frac{r^{-N}}{2\pi \gamma (1+a^{2})}\frac{%
\left[ 1+\frac{r_{+}(a^{2}-1)}{2r}\right] }{\sqrt{1-\frac{r_{+}}{r}}},
\end{equation}

The meaning of this result is that an accelerating observer, who sits on the
AS with $r$ distance far away from the center of the CDBH always detects
thermal radiation with $T_{U}.$ In other words, from the viewpoint of the
accelerating observer, the space of the inertial observer in which it is
filled with the quantized fields that make up a universe around CDBH looks
like a state containing many particles in thermal equilibrium with $T_{U}$.
It is noteworthy that the Unruh effect emerges from the local quantum fields 
\cite{Qfields}. In the Schwarzschild limit ($a^{2}\rightarrow \infty $ with $%
\gamma =1$ ), the Unruh temperature (22) reduces to

\begin{equation}
T_{U}\rightarrow T_{U_{Schw.}}=\frac{M}{2\pi r^{2}}\frac{1}{\sqrt{1-\frac{2M%
}{r}}},
\end{equation}

which has the Newtonian limit at large distances $r\gg $

\begin{equation}
\lim_{r\rightarrow \infty }T_{U_{Schw.}}=\frac{M}{2\pi r^{2}},
\end{equation}

Since our aim is to attain the entropic force on the AS of the CDBH, the
flux $\Phi _{AS}$\ of the proper acceleration through the AS should be taken
into account. To this end, we use the definition of $\Phi _{AS}$ given by 
\cite{Makela}

\begin{equation}
\Phi _{AS}=4\pi \tilde{a}R^{2},
\end{equation}

from which it readily follows that

\begin{equation}
\Phi _{AS}=\frac{4\pi \gamma }{1+a^{2}}r^{-2N}\frac{\left[ 1+\frac{%
r_{+}(a^{2}-1)}{2r}\right] }{\sqrt{1-\frac{r_{+}}{r}}},
\end{equation}

Introducing the change of heat $\delta Q$ as

\begin{equation}
\delta Q=\frac{1}{4\pi }d\Phi _{AS},
\end{equation}

where $d\Phi _{AS}$ denotes the differential of the flux $\Phi _{AS}$\
through the AS. Let us suppose that the proper acceleration $\tilde{a}$ is
uniform on the AS when varying the flux $\Phi _{AS}.$ This corresponds to $d%
\tilde{a}(r,M)=0$ such that the following relation automatically appears:

\begin{equation}
dM=\frac{\{(a^{4}-1)M\left[ \left( 1+a^{2}\right) \left( 3a^{2}+1\right)
M-2a^{4}r\right] -a^{6}r^{2}\}}{r(1+a^{2})^{2}\left[ (a^{4}-1)M-a^{4}r\right]
}dr,
\end{equation}

Inserting this into Eq. (27), and making some algebra, we end up with the
following equation

\begin{equation}
\delta Q=\frac{2\gamma N}{1+a^{2}}r^{-\frac{1}{1+a^{2}}}\frac{\left[ 1+\frac{%
r_{+}(a^{2}-1)}{2r}\right] }{\sqrt{1-\frac{r_{+}}{r}}}dr,
\end{equation}

We may define the entropy as

\begin{equation}
S_{AS}=2\pi \gamma ^{2}r^{2N},
\end{equation}

whence its differential can be calculated as

\begin{equation}
dS_{AS}=4\pi \gamma ^{2}Nr^{\frac{a^{2}-1}{a^{2}+1}}dr,
\end{equation}

Plugging the above equation into Eq. (29),

\begin{eqnarray}
\delta Q &=&\frac{r^{-N}}{2\pi \gamma (1+a^{2})}\frac{\left[ 1+\frac{%
r_{+}(a^{2}-1)}{2r}\right] }{\sqrt{1-\frac{r_{+}}{r}}}dS_{AS}  \notag \\
&=&T_{U}dS_{AS}=T_{AS}dS_{AS},
\end{eqnarray}

This result represents a fact that the change of heat is balanced by the
change of the entropy when fixing the temperature on the AS.

Finally, the entropic force, from Eq.(3), is obtained as,

\begin{equation}
F_{AS}=\frac{m}{\gamma \left( 1+a^{2}\right) }r^{-N}\frac{\left[ 1+\frac{%
r_{+}(a^{2}-1)}{2r}\right] }{\sqrt{1-\frac{r_{+}}{r}}},
\end{equation}

which is exerting on the object with mass $m$ near to the AS of the CDBH
having mass $M$. This result shows us that any object possessing mass $m$
will be under the influence of infinitely tidal force while it is close to
the horizon, $r\rightarrow r_{+}.$ At large distance limit, $r>>r_{+}$, $%
F_{AS}$ takes the following form

\begin{equation}
F_{AS}\rightarrow F_{MN}=\frac{m}{\gamma }r^{-N}(\frac{1}{1+a^{2}}+\frac{M}{r%
}).
\end{equation}

One can easily see that the above equation takes the Newtonian force law (8)
if $a^{2}\rightarrow \infty ,$ and $\gamma =1$\ is chosen. That is why, we
prefer to symbolize the large distance limit of $F_{AS}$ as $F_{MN}$, and
designate it as modified Newtonian force. As illustrated in Fig.1, the value
of the dilaton parameter $a$ is effective on $F_{MN},\ $which is exerted by
the mass $M$ of the CDBH on the object possessing mass $m$. Here, we
considered a small object compared to the mass of the BH. It is clear that
when $a\rightarrow 0,$ $F_{MN}$ increases. This intriguing result might also
be used to prove the existence of the dilaton fields in the future. On the
other hand, it is observed that when $a$ gets higher values than zero, up to
infinity, $F_{MN}$ behaves like the conventional Newtonian force. In the
caption of Fig. 1, the values of the physical parameters are depicted.

\section{Static Holographic Screen of CDBH and the Entropic Force}

In this section, we briefly review how one derives the entropic force on the
static HS, which belongs to the CDBH (10). For this purpose, we first
consider the following definition for nonzero component of the proper
acceleration vector \cite{Tian}

\begin{equation}
a^{\mu }=\frac{1}{f}\xi ^{\alpha }\nabla _{\alpha }\xi ^{\mu },
\end{equation}

where $\xi ^{\upsilon }$ is a timelike Killing vector. In order to define
the concept of energy in the stationary spacetime, one might use 
\begin{equation}
E(V)=\frac{1}{4\pi }\doint a^{\mu }n_{\mu }dA,
\end{equation}

where $V$ is the bulk volume enclosed by spacelike hypersurface $S^{2}$ i.e. 
$\partial V=S^{2}$. In general, the nonzero component of the Killing vector
for the outer region of the CDBH ($r>r_{+}$) is given by $\xi ^{\alpha
}=[1,0,0,0].$ By using these quantities, we can easily evaluate the integral
(36) and obtain the energy, which is defined on the HS of the CDBH as follows

\begin{equation}
E_{HS}=\frac{\gamma }{1+a^{2}}r^{2-N}\frac{\left[ 1+\frac{r_{+}(a^{2}-1)}{2r}%
\right] }{\sqrt{1-\frac{r_{+}}{r}}},
\end{equation}

One can easily verify that while $a^{2}\rightarrow \infty $ with $\gamma =1,$
it reduces to

\begin{equation}
E_{HS}\rightarrow \frac{M}{\sqrt{1-\frac{r_{+}}{r}}},
\end{equation}

which is the energy of the HS for the Schwarzschild BH. Namely, our result
(37) is consistent with the HS energy result reported in \cite{Myung}. When
an observer is at rest and located at coordinate $r$ with respect to CDBH
spacetime, $E_{HS}$ corresponds to the energy of the gravitational field,
which is measured by that observer. After imposing the local equipartition
rule

\begin{equation}
E_{HS}=2S_{HS}T_{HS},
\end{equation}

where $S_{HS}$ indicates the entropy on the HS located at $r,$ where is
outside of the horizon. $S_{HS}$ is found by

\begin{equation}
S_{HS}=\pi R^{2},
\end{equation}

Thus, by using of Eq. (39) one can obtain the temperature on the HS of CDBH
as

\begin{equation}
T_{HS}=\frac{1}{2\pi \gamma \left( 1+a^{2}\right) }r^{-N}\frac{\left[ 1+%
\frac{r_{+}(a^{2}-1)}{2r}\right] }{\sqrt{1-\frac{r_{+}}{r}}},
\end{equation}

Considering Eq.(32), we see that $T_{HS}=T_{AS}$. Namely, the bulk
temperature $T_{AS}$ is exactly equal to the HS temperature $T_{HS}.$

Finally, inserting Eq.(41) into Eq. (3), we attain the entropic force on the
object $m$ near the HS as

\begin{equation}
F_{HS}=2\pi mT_{HS}=\frac{m}{\gamma \left( 1+a^{2}\right) }r^{-N}\frac{\left[
1+\frac{r_{+}(a^{2}-1)}{2r}\right] }{\sqrt{1-\frac{r_{+}}{r}}},
\end{equation}

which means that the entropic force (33) is recovered impeccably: $F_{HS}=$ $%
F_{AS}.$

As a final remark for this section, we would like to highlight that
temperature $T_{HS}$, which belongs to an spacetime (10) with unusual
asymptotics is obtained by the using of the local equipartition rule.
Thence, the entropic force (3) is read for those unusual asymptotical CDBHs.
All deserved limits of that entropic force (42) can be easily verified.
Consequently, the procedure followed up to obtain the entropic force shows
us that it is not only applicable to AF BHs, but it could be used for NAF
BHs as well.

\section{Stretched Horizon of CDBH and the Entropic Force}

In this section, the SH is considered as a HS, and we explore the form of
the entropic force on the SH. The concept of SH \cite{SHorizon} arouses
interest because of its location. Its location is very close to the event
horizon (near horizon limit of the CDBH or the Rindler space \cite{Rindler}
of the CDBH), and all temperatures on the SH that are obtained by different
approaches become equivalent to each other. Through the section, the radial
distance of the SH from the center of the CDBH is considered as $%
r=r_{+}+1/r_{+}$ in which $r_{+}\gg 1$. Briefly, the SH is located just
above the event horizon.

The AS and its equivalent HS temperatures, which are represented by Eqs.
(22) and (41) take the following forms on the SH.

\begin{eqnarray}
T_{AS}^{SH} &=&T_{HS}^{SH}=\frac{1}{4\pi \gamma }\left( r_{+}\right) ^{\frac{%
1}{1+a^{2}}}(1+\frac{1}{r_{+}^{2}})^{-\frac{3a^{2}+1}{2\left( 1+a^{2}\right) 
}}\left[ 1+\frac{2}{r_{+}^{2}(1+a^{2})}\right]  \notag \\
&\simeq &\frac{1}{4\pi \gamma }\left( r_{+}\right) ^{\frac{1}{1+a^{2}}}\left[
1-\frac{3\left( a^{2}-1\right) }{2r_{+}^{2}\left( 1+a^{2}\right) }\right] ,
\end{eqnarray}

Since the near horizon limit of a BH is also known as the Rindler space of
the considered BH, the geometry of a SH can be described by a Rindler
spacetime. Besides, the local Rindler temperature \cite{LocRindler} is given
by

\begin{equation}
T_{LR}^{SH}=\left. \frac{\kappa (r)}{2\pi \sqrt{-g_{00}}}\right\vert
_{r=r_{+}+1/r_{+}}=\left. \frac{f^{\prime }}{4\pi \sqrt{f}}\right\vert
_{r=r_{+}+1/r_{+}},
\end{equation}

which yields

\begin{eqnarray}
T_{LR}^{SH} &=&\frac{1}{2\pi \gamma (1+a^{2})}\left( r_{+}\right) ^{\frac{1}{%
1+a^{2}}}\left[ \frac{1}{2}(1+a^{2})+\frac{1}{r_{+}^{2}}\right] ,  \notag \\
&\simeq &\frac{1}{4\pi \gamma }\left( r_{+}\right) ^{\frac{1}{1+a^{2}}},
\end{eqnarray}

As envisioned before, all temperatures on the SH, which are obtained by
using different approaches are equal to each other in the leading order:

\begin{equation}
T_{AS}^{SH}=T_{HS}^{SH}=T_{LR}^{SH}=\frac{1}{4\pi \gamma }\left(
r_{+}\right) ^{\frac{1}{1+a^{2}}},
\end{equation}

As it can be seen from the above equation, unless $a\rightarrow \infty ,$
the SH temperature will depend on the quasilocal mass $M.$

On the other, the definition of the local energy is given by

\begin{equation}
E_{L}=\frac{M}{\sqrt{-g_{00}}},
\end{equation}

In the Rindler space, the foregoing takes the following form

\begin{eqnarray}
E_{LR}^{SH} &\simeq &\gamma M\left( r_{+}\right) ^{N}  \notag \\
&=&\frac{NA_{EH}}{8\pi \gamma }\left( r_{+}\right) ^{\frac{1}{1+a^{2}}}
\end{eqnarray}

It is possible to obtain the same result when Eq.\ (37) is used for in
defining the local energy of the holographic screen on the stretched horizon
-- $E_{LR}^{SH}=E_{HS}^{SH}.$ Significantly, one can easily verify that the
equipartition of energy $E_{HS}^{SH}=\frac{1}{2}\tilde{N}^{SH}T_{HS}^{SH}$
is satisfied. However, the number of bits stored on the holographic screen
is found as $\tilde{N}^{SH}=NA_{EH}.$ When comparing this result with Eq.
(5), we see that the proportionality constant $c=N,$ which changes between $%
0^{+}$ and $1$ i.e., \ $0^{+}<c<1$ depending on the value of the dilaton
parameter $a$. In other words, the dilaton parameter $a$ scales the total
information stored on the screen such that it reaches its maximum when $%
a\rightarrow \infty $, which corresponds to the vanishing of the dilaton
field or to the AF\ spacetime. In terms of the Bekenstein-Hawking entropy

\begin{equation}
S_{BH}=\frac{A_{EH}}{4},
\end{equation}

one can easily see that $\tilde{N}^{SH}=4NS_{BH}$. After imposing the first
law of thermodynamics

\begin{equation}
dE_{HS}^{SH}=T_{HS}^{SH}dS^{SH}
\end{equation}

we see that the entropy on the SH should be

\begin{equation}
S^{SH}=\frac{1+2a^{2}}{1+a^{2}}S_{BH}
\end{equation}

Obviously, the factor $\frac{1+2a^{2}}{1+a^{2}}$\ always prevents the
equality of both entropies. Nevertheless, in the case of ultrahigh dilaton
field, corresponding to the limit of $a\rightarrow 0^{+}$, $S^{SH}$ and $%
S_{BH}$ become almost same. Conversely, while $a\rightarrow \infty ,$ $%
S^{SH} $ becomes double of $S_{BH},$ as can be seen in \cite{Myung}. Hence,
one can remark that depending on the dilaton parameter $a$, the rate $\frac{%
S^{SH}}{S_{BH}}$ changes in the manner that $1<\frac{S^{SH}}{S_{BH}}<2$.

Finally, according to our established entropic force derivation, it is now
necessary to substitute Eq. (46) into Eq. (3). Thus, the entropic force on
the SH yields

\begin{eqnarray}
F_{SH} &=&2\pi m(T_{AS}^{SH}=T_{HS}^{SH}=T_{LR}^{SH})  \notag \\
&=&\frac{1}{2\gamma }m\left( r_{+}\right) ^{\frac{1}{1+a^{2}}}  \notag \\
&=&m\tilde{a}_{SH}
\end{eqnarray}

where $\tilde{a}_{SH}=\frac{1}{2\gamma }\left( r_{+}\right) ^{\frac{1}{%
1+a^{2}}}$\ is the proper acceleration on the SH. Alternatively, $\tilde{a}%
_{SH}$ can alternatively be obtained by taking the near horizon limit ($%
r=r_{+}+1/r_{+}$) of Eq. (21).

\section{Conclusion}

After making the remarkable connections between gravity and thermodynamics,
we showed that whenever the temperature on a holographic screen is read
precisely, it is easy to calculate the entropic force by using $F=2\pi mT$.
In addition to this, it is also shown that Verlinde's entropic force
formalism is applicable to the CDBHs, which are neither AF nor NAF
spacetimes. As a result, we support the Verlinde's arguments, which imply
that gravity is not a fundamental force but emergent.

In this paper, we introduced three possible distinct surfaces (or screens),
which are separately being a candidate for the HS of a CDBH. Those are the
AS, the static HS and the SH or the Rindler spacetime of the CDBH. Among
them, the SH is the special one. Because, the SH is located at such a
specific place that when all the other obtained temperatures are
recalculated for the location of the SH, they all, and thus their
corresponding entropic forces, become equivalent. On the other hand, the
entropic forces obtained by using of the AS and of the HS reduce to the
entropic force of the Schwarzschild BH when the dilaton field is terminated,
and subsequently to the Newton's force law at large distances. Dependence of
the entropic force of a small object (compared with the mass $M$ of the BH)
having mass $m$ on the dilaton parameter at large distances is graphically
illustrated. The graph shows us that a distant observer may feel the gravity
of the CDBH higher than the Newton's force if the strength of the dilaton
field increases. The latter result might provide a clue to experimentally
test the concept of entropic force, or even to detect the dilaton fields.

Finally, we should say that the discussions on the Verlinde's arguments have
been continuing without pausing, even day by day one can see many new
studies, which are referring the Verlinde's work \cite{Verlinde} pro and con
(see for instance \cite{procon}). Although we sustain that our analysis is
not adequate to fully understand the entropic force, here our ultimate aim
is to make a contribution to the literature that the derivation of the
entropic force according to the Verlinde's arguments is also possible in the
spacetimes with unusual asymptotics. Moreover, we would like to keep the
topic fresh. Because we believe that even the Verlinde's arguments turn out
to be in the wrong, hashing the topic of the entropic force over with other
researchers working in the same field will guide us for better understanding
the origin of gravity.

\begin{acknowledgments}
I am grateful to Professor M. Halilsoy for fruitful discussions and
suggestion of studying on the theory of emergent gravity.
\end{acknowledgments}

\bigskip

\textbf{Figure Caption:}

Figure 1: Entropic Force $F_{MN}$ versus the dilaton parameter $a$ for the
CDBHs. The plot is governed by Eq. (34). The physical parameters in Eq. (34)
are chosen as follows: $\gamma =1$, $M=1$ and $m=1\times 10^{-6}.$

\bigskip 

\end{document}